 \documentclass[aps,prx,twocolumn,showpacs,superscriptaddress,longbibliography]{revtex4-2}
\usepackage{graphicx}
\usepackage{epstopdf}
\usepackage{float}
\usepackage{color}
\usepackage{amsmath}
\usepackage{bm}

\begin{document}

\title{Slow ferromagnetic fluctuations in the kagome metal Sc$_3$Mn$_3$Al$_7$Si$_5$ revealed by $^{27}$Al NMR}

\author{Qing-Ping Ding}
\affiliation{Ames National Laboratory, U.S. DOE, Ames, Iowa 50011, USA}

\author{Charles Taylor}
\affiliation{Ames National Laboratory, U.S. DOE, Ames, Iowa 50011, USA}
\affiliation{Department of Physics and Astronomy, Iowa State University, Ames, Iowa 50011, USA}

\author{Yongbin Lee}
\affiliation{Ames National Laboratory, U.S. DOE, Ames, Iowa 50011, USA}

\author{Charuni Dissanayake}
\email{Present address: National High Magnetic Field Laboratory, Florida State University, Tallahassee, FL 32310}
\affiliation{Department of Physics, University of Central Florida, Orlando, Florida 32816, USA}

\author{Vireshwar Mishra}
\email{Present address: Department of Physics, Morgan State University, Richard N. Dixon Science Research Center, Baltimore, Maryland 21251, USA}
\affiliation{Department of Physics, University of South Florida, Tampa, Florida 33620, USA}

\author{Dang Khoa Le}
\affiliation{Department of Physics, University of South Florida, Tampa, Florida 33620, USA}

\author{Manh-Huong Phan}
\affiliation{Department of Physics, University of South Florida, Tampa, Florida 33620, USA}

\author{Yasuyuki Nakajima}
\affiliation{Department of Physics, University of Central Florida, Orlando, Florida 32816, USA}

\author{Yuji Furukawa}
\affiliation{Ames National Laboratory, U.S. DOE, Ames, Iowa 50011, USA}
\affiliation{Department of Physics and Astronomy, Iowa State University, Ames, Iowa 50011, USA}

\date{\today}

\begin{abstract}
     Static and dynamical magnetic and electronic properties of the kagome metal Sc$_3$Mn$_3$Al$_7$Si$_5$  have been investigated by $^{27}$Al nuclear magnetic resonance (NMR) measurements.      The temperature dependence of  Knight shift ($K$) shows a similar temperature dependence of the DC  magnetic susceptibility $\chi$ except for the low-temperature region below $\sim$ 50 K where  $K$ is almost constant while $\chi$ keeps increasing, which suggests that the increase in $\chi$ at low temperatures is not intrinsic.      $^{27}$Al spin-lattice relaxation rate divided by temperature ($1/T_1T$)  is found to be constant,  confirming the metallic state of Sc$_3$Mn$_3$Al$_7$Si$_5$ from a microscopic point of view.           Based on a Korringa ratio analysis using the $T_1$ and  $K$  data,  ferromagnetic spin fluctuations are found to dominate in Sc$_3$Mn$_3$Al$_7$Si$_5$.  These fluctuations are suggested to be very slow with frequencies on the order of kilohertz or lower.
  \end{abstract}

\maketitle
Recently there has been growing interest in metals with kagome lattice structures, primarily due to possible topological nontrivial electronic states originating from flat band structures \cite{Yin2022}.
   $RM_6$Sn$_6$   system where  $R$ represents rare earth elements and $M$ refers to 3$d$ transition metals  (so-called 166-kagome metals) is one of the relatively well-studied kagome metals \cite{Xu2023}. 
   In Mn-based 166 compounds, spin-polarized Dirac cones, anomalous Hall effects, and Chern topological magnetism have been pointed out in its ferrimagnetic state of TbMn$_6$Sn$_6$ \cite{Yin2020,Riberolles2022,Riberolles2023,Riberolles2024}. 
   The dispersionless electronic edge states within the gapped Dirac cones are also suggested, especially when spin-orbit coupling and out-of-plane ferromagnetism are considered \cite{Riberolles2024}.  
  With non-magnetic V ions replaced for magnetic Mn ions,  a charge density wave (CDW)  phase has been reported in ScV$_6$Sn$_6$ \cite{Cao2023} while a ferromagnetic order state appears below 5 K in GdV$_6$Sn$_6$ \cite{Pokharel2021}. 
  Intriguing phenomena such as superconductivity, CDWs, pair density waves, and electronic nematicity, have also been observed in another family of kagome metals of $A$V$_3$Sb$_5$ ($A$ = K, Rb, Cs) \cite{Ortiz2019,Ortiz2020,Wilson2024}. 
These findings make kagome metals a versatile platform for exploring a wide variety of intriguing physical properties.

 \begin{figure}[b]
\includegraphics[width=0.9\columnwidth]{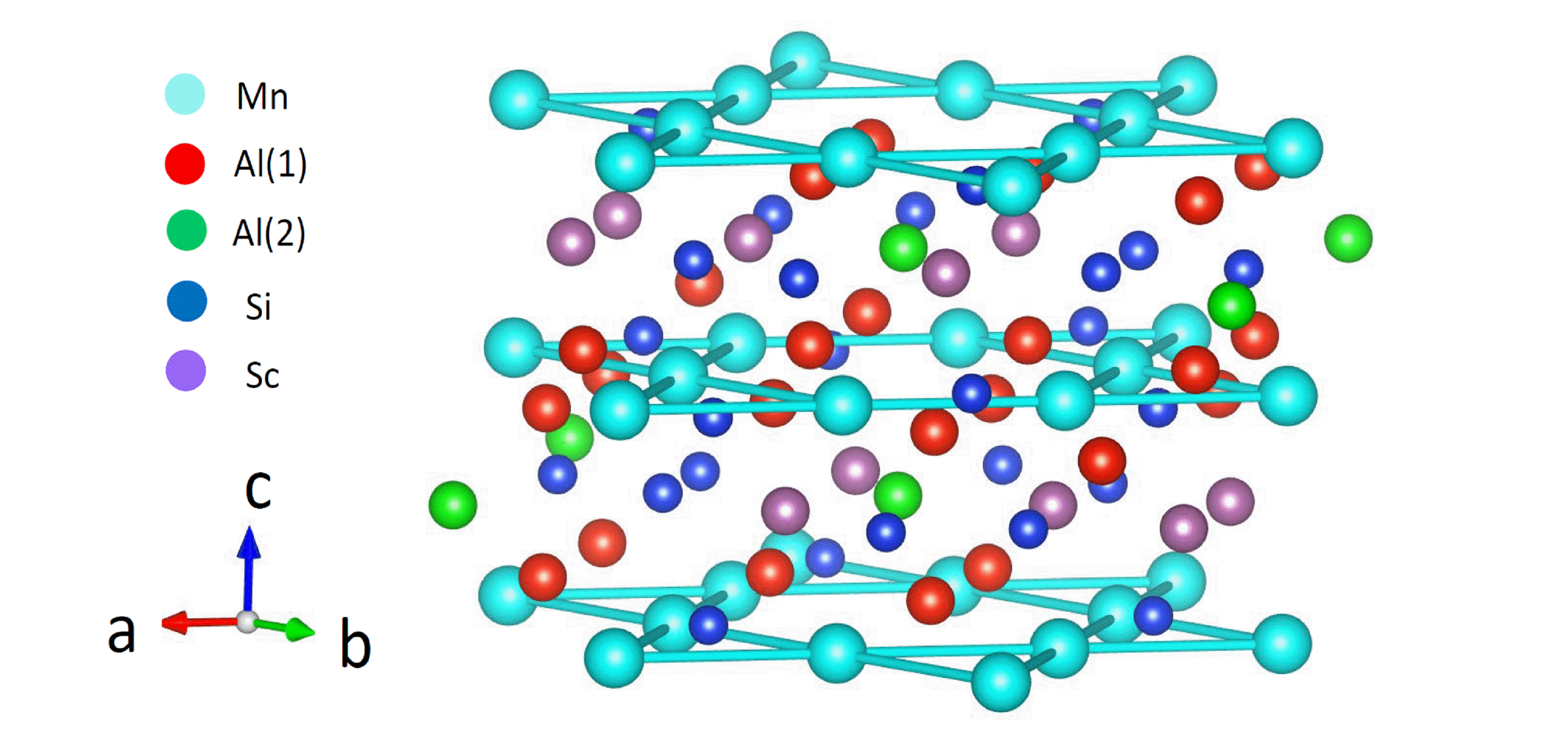} 
\caption{Schematic view of the crystal structure of Sc$_3$Mn$_3$Al$_7$Si$_5$. Note Si atoms are represented by blue circles without showing the two different Si sites, 6h and 4f.}
\label{fig:structure}
\end{figure}

  Recently, a Mn-based kagome metal Sc$_3$Mn$_3$Al$_7$Si$_5$ has been reported to be located close to a ferromagnetic instability originating from the flat band, one of the characteristic properties of kagome metals  \cite{Samanta2024}.
   Sc$_3$Mn$_3$Al$_7$Si$_5$ crystallizes in a hexagonal structure (space group: P6$_3/mmc$) with the lattice parameters of $a$ = $b$ = 8.3519(2)~\AA~ and $c$ = 9.0845(4)~\AA~at $T$ = 299 K \cite{He2014}. 
   It is one of the rare examples of metallic kagome systems where Mn ions form two-dimensional kagome layers separated by Si and Sc ions as shown Fig.  \ \ref{fig:structure}.
There are six unique crystallographic sites: Sc (6h), Mn (6g), Al1 (12k), Al2 (2b), Si1 (6h), and Si2 (4f) with the numbers in parentheses representing the Wyckoff position of each atom. 
       The metallic nature of the compound has been confirmed through its temperature ($T$) dependence of resistivity. 
       No magnetic order has be observed down to 1.8 K \cite{He2014}.
      A relatively large Sommerfeld coefficient with 57-80 mJ/mol K$^2$ \cite{He2014, Li2021} has been reported from the specific heat measurements,  suggesting strong electron correlations.
     Initially, static magnetic susceptibility $\chi$ was reported to be isotropic and follows Curie-Weiss (CW) law above $\sim$ 50 K  with an effective magnetic moment $\mu_{\it eff}$ of 0.51 $\mu_{\it B}$/Mn and a negative Weiss temperature $\theta$ of -38  K \cite{He2014}. 
    On the other hand, a more recent study by Samanta et al. reported slightly anisotropic $\chi$ and relatively different values of $\mu_{\it eff}$ of 0.86 (0.87) $\mu_{\it B}$/Mn and $\theta$ of -421 (-369) K for magnetic field $H$ parallel to the $ab$ plane (the $c$ axis) \cite{Samanta2024}. 
   In addition, despite the antiferromagnetic negative Weiss temperatures, Samanta et al. suggested the ferromagnetic fluctuations from the increase in $\chi$ at low temperatures, as well as $^{27}$Al nuclear magnetic resonance (NMR), transport and optical conductivity measurements \cite{Samanta2024}.

 \begin{figure*}[h!tb]
\includegraphics[width=2.1\columnwidth]{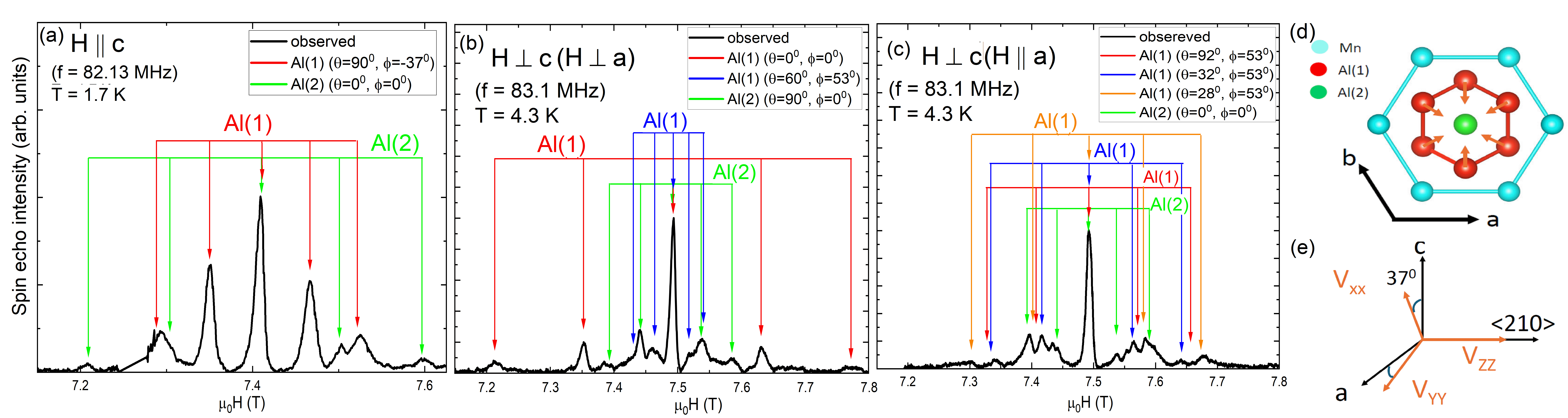}
\caption{(a) $H$-swept $^{27}$Al-NMR spectrum for $H \parallel c$ measured at 1.7 K and $f$ = 82.13 MHz.  The red and green lines are the positions of $^{27}$Al-NMR spectra calculated with  $\nu_{\rm Q}$~=1.55(2)~MHz, $\eta$ = 0.66, $\theta$ = 90$^{\circ}$, and $\phi$ = 37$^{\circ}$ for Al(1) and $\nu_{\rm Q}$~=1.07(2)~MHz, $\eta$ = 0, $\theta$ = 0$^{\circ}$, and $\phi$ = 0$^{\circ}$ for Al(2), respectively.  (b) $H$-swept $^{27}$Al-NMR spectrum for $H\perp c$ (and also $H\perp a$) at 4.3 K and $f$ = 83.1 MHz. The red (blue) lines are the calculated positions of Al(1) NMR spectra with  $\theta$ = 0$^{\circ}$ (60$^{\circ}$) and $\phi$ = 0$^{\circ}$ (53)$^{\circ}$, with the same values of  $\nu_{\rm Q}$~=1.55(2)~MHz, $\eta$ = 0.66.   (c)  $H$-swept $^{27}$Al-NMR spectrum for $H \perp c$ (and also $H\parallel a$) at 4.3 K and $f$ = 83.1 MHz.   The green lines in (b) and (c) represent the positions of NMR lines for Al(2) with $\nu_{\rm Q}$~=1.07(2)~MHz, $\eta$ = 0, $\theta$ = 90$^{\circ}$, and $\phi$ = 0$^{\circ}$.  The other lines in (c) are simulated spectra (see the text for details). (d)  Top view of the kagome constructed with Mn ions (cyan) with Al(1) (red) and Al(2) (green). The arrows on Al(1) represent the principal axis of $V_{\rm ZZ}$ for three inequivalent Al(1) sites.  (e)  The schematic view of the directions for the principal axes of the EFG with respect to the crystalline axes determined by the NMR measurements and the electronic band structure calculation. }
\label{fig:spectrum}
\end{figure*}

    In this paper, we report the detailed $^{27}$Al NMR studies on Sc$_3$Mn$_3$Al$_7$Si$_5$. 
    Two distinct $^{27}$Al NMR  signals originating from  the Al(1) and Al(2) sites are observed.  
     Based on the results of Knight shift ($K$),  nuclear spin-lattice relaxation rate (1/$T_1$) and nuclear spin-spin relaxation rate (1/$T_2$) of the Al(1) site, very slow ferromagnetic spin fluctuations are found to dominate in Sc$_3$Mn$_3$Al$_7$Si$_5$ at low temperatures. 
     Our work suggests that such slow ferromagnetic fluctuations play an important role in the peculiar magnetic and electronic states in Sc$_3$Mn$_3$Al$_7$Si$_5$.

      Rod-like single crystals (with hexagonal-shaped cross-section) of Sc$_3$Mn$_3$Al$_7$Si$_5$  were grown out of high temperature solutions \cite{He2014}.
       The crystalline $c$ axis and the $ab$ plane are parallel and perpendicular to the rod direction of the crystal, respectively. 
      NMR  measurements of $^{27}$Al ($I$ = $\frac{5}{2}$, $\frac{\gamma_{\rm N}}{2\pi}$ = 11.0943 MHz/T, $Q=$ 0.149 barns) nuclei were conducted using a laboratory-built phase-coherent spin-echo pulse spectrometer. 
       The NMR spectra were obtained by sweeping $H$ at fixed frequencies. 
      $H$ was applied parallel to either the crystalline $c$ axis ($H \parallel c$) or the $ab$ plane ($H$ $\parallel$ $ab$).
      For the NMR spectrum measurements at the lowest $T$ of 1.7 K for $H$ $\parallel$ $c$ and 4.3 K for   $H$ $\parallel$ $ab$, we use only one single crystal.
     For high-$T$ measurements, we used several aligned crystals to increase the signal intensity.
      The $^{27}$Al 1/$T_{\rm 1}$ was measured with a saturation recovery method.
   $1/T_1$ at each $T$ was determined by fitting the nuclear magnetization $M$ versus time $t$  using the exponential function $1-M(t)/M(\infty) = 0.028e^{-t/T_1}+0.178e^{-6t/T_1}+0.794e^{-15t/T_1}$,  where $M(t)$ and $M(\infty)$ are the nuclear magnetization at time $t$ after the saturation and the equilibrium nuclear magnetization at $t$ $\rightarrow$ $\infty$, respectively, for the case of magnetic relaxation \cite{Recovery}. 
     All the nuclear magnetization recovery curves measured were well fitted with the function.
    Nuclear spin-spin relaxation time $T_2$ was determined by fitting spin-echo decay curves using  the following equation:
$M(2\tau)= M(0)~$exp$[-(\frac{2\tau}{T_2})^\beta$], 
where $\tau $ is the time between $\pi$/2 and $\pi$ pulses.
Above 100 K,  Gaussian decay behavior with  $\beta \approx 2$ was observed. 
With decreasing $T$,  $\beta$ decreases below 100 K and becomes  $\approx 1$ around 7 K and then increases to $\sim$ 1.5 at lower $T$.
        For the analysis of the NMR data, we measured the magnetic susceptibility $\chi(T)$ of the single crystal at $\mu_0H$ = 7.4 T applied parallel to the $c$ axis and to the $ab$ plane in a commercial Quantum Design Physical Property Measurement System (PPMS)  using a Vibrating Sample Magnetometer, and we found that the observed $\chi(T)$ curves were nearly isotropic, consistent with the previous report \cite{He2014}.


   Figure  \ \ref{fig:spectrum}(a)  shows the $H$-swept $^{27}$Al NMR spectrum for $H \parallel c$ axis at 1.7 K where two sets of $^{27}$Al NMR lines were observed with different intensity.
   The typical spectrum for a nucleus with spin $I=5/2$ with Zeeman and quadrupolar interactions can be described by the following nuclear spin Hamiltonian \cite{Slichter_book} which produces a spectrum with a central transition line flanked by two satellite lines on both sides.  
\begin{eqnarray}
\centering
{\cal H} &=& -\gamma_{\rm N}\hbar(1+K) {\bf H} \cdot {\bf I} + \frac{h\nu_{\rm Q}}{6}(3I_Z^2-I^2  +\frac{1}{2}\eta(I_+^2 +I_-^2)),
\label{eq:1}
\end{eqnarray} 
where $H$ is external field, $\hbar$ is Planck's constant ($h$) divided by 2$\pi$,  and $K$ represents the NMR shift.  
The first and second terms represent Zeeman and quadrupolar interactions, respectively.  
   The nuclear quadrupole frequency for $I=5/2$ nuclei is given by $\nu_{\rm Q} = 3eQV_{\rm ZZ}/20h$, where $Q$ is the nuclear quadrupole moment and $V_{\rm ZZ}$ is the maximum electric field gradient (EFG) at nuclear sites.
  $\eta$ is the asymmetry parameter of EFG  defined by  $\frac{V_{\rm XX} -V_{\rm YY}}{V_{\rm ZZ}}$ with $|V_{\rm ZZ}|$$\geq$$|V_{\rm YY}|$$\geq$$|V_{\rm XX}|$. 
    In this case, the resonance frequency $f$ for the transition from $I_z$ = $m$ to $m-1$ is given within first-order perturbation theory by \cite{Abragambook}
 \begin{eqnarray*}
 \lefteqn{ f(m \leftrightarrow m-1)}  \\
& =&   f_0 + \frac{1}{2}\nu_{\rm Q}(m-\frac{1}{2})(3\cos^2\theta-1+\eta \sin^2\theta \cos2\phi ). 
  \label{eq:3}
  \end{eqnarray*} 
       Here $ f_0 = \frac{\gamma_{\rm N}}{2 \pi}H$, and   $\theta$ and $\phi$ are the polar and azimuthal angles between the $Z$ axis of the EFG and the direction of $H$, respectively, where the quantization axis ($z$ axis) for the Zeeman interaction points along the $H$ direction. 
    Thus, the NMR spectra, especially the spacing between the lines, depend on the angles $\theta$ and $\phi$ as well as $\nu_{\rm Q}$ and $\eta$. 
   As mentioned above, there are two Al sites [Al(1) and Al(2)]  and the local symmetries at Al(1) and Al(2) are .m. and $\bar{6}$m2, respectively,  in Sc$_3$Mn$_3$Al$_7$Si$_5$ \cite{He2014}. 
   This means that there is a 6 rotoinversion around the $c$ axis for the Al(2) site, which results in  $\eta$ = 0  and $V_{\rm ZZ}$ is parallel to the $c$ axis for Al(2). 
     In contrast, as the local symmetry of Al(1) is .m., the direction of the principal axes of EFG is not trivial. 
     Therefore,   to obtain more information about the values of $\eta$ and $\nu_{\rm Q}$, and the principal axes of the EFG at the Al(1) site as well as the  $\nu_{\rm Q}$ value for Al(2),   we have calculated the EFG at each Al site by a full potential linear augmented plane wave (FLAPW) method \cite{FLAPW} with a generalized gradient approximation \cite{GGA} using the lattice parameters described in Introduction.
    The $\nu_{\rm Q}$ values were calculated to be 1.56 MHz and 1.01 MHz for Al(1) and Al(2), respectively.  
    We also found from the calculations that $V_{\rm ZZ}$ is parallel to the $c$ axis and the $\eta$ is zero for Al(2) as expected from the local symmetries of the Al(2) site. 
   As for the Al(1) site, $V_{ZZ}$ is calculated to be parallel to $<$210$>$ (i.e., perpendicular to the $a$ axis)  with the $\eta$ = 0.66 and the principle axis of $V_{\rm XX}$ is rotated by 37$^{\circ}$ from the $c$ axis in the $ac$ plane  as shown in  Fig.  \ \ref{fig:spectrum}(e). 
 
 \begin{figure}[tb]
\includegraphics[width=0.95\columnwidth]{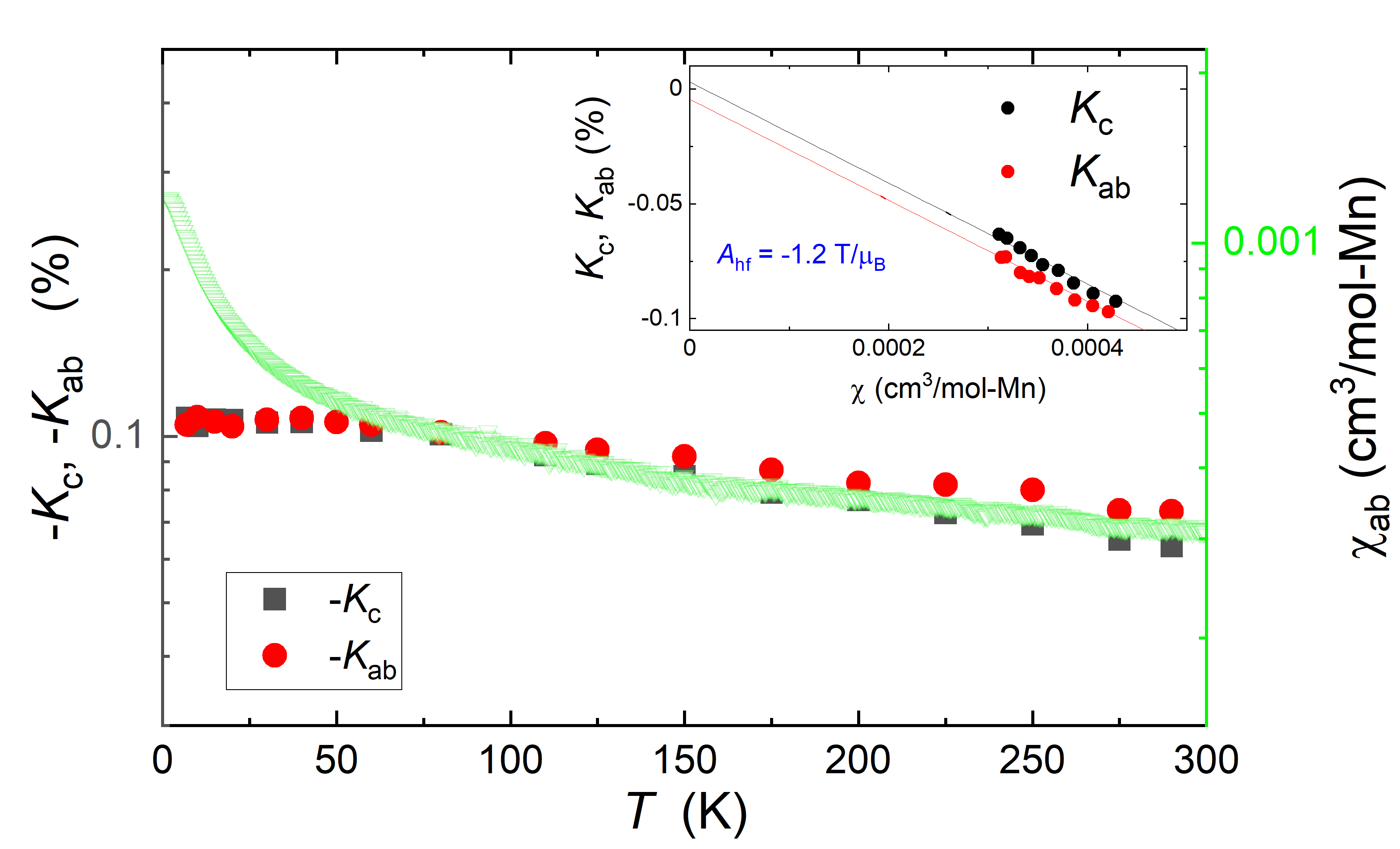} 
\caption{$T$ dependence of Knight shifts $K_c$  and $K_{ab}$ together with the $T$ dependence of $\chi_{ab}$ (green circles) measured at 7.4 T. 
Note that Knight shift data are plotted with $-K_c$  and $-K_{ab}$ in logarithmic scale.  The inset shows the $K$-$\chi$ plot for $K_c$ and $K_{ab}$ versus the corresponding $\chi$.}
\label{fig:Shift}
\end{figure}

  First, we consider the signal intensity to assign the lines. 
As the Wyckoff positions are 12k and 2b for Al(1) and Al(2), respectively,  the NMR signal intensity from Al(1) is considered to be six times as large as that of Al(2). 
   Thus the set of NMR lines with lower intensities are assigned to Al(2) and are well explained by a set of parameters of $\nu_{\rm Q}$~=1.07(2)~MHz, $\eta$ = 0, $\theta$ = 0$^{\circ}$, and $\phi$ = 0$^{\circ}$ as shown by the green lines in Fig.  \ \ref{fig:spectrum}(a), in a very good agreement with   $\nu_{\rm Q}$~=1.01~MHz from the calculations.
   As for Al(1),  we also find the observed NMR lines can be reproduced well with the parameters $\nu_{\rm Q}$~=1.55(2)~MHz, $\eta$ = 0.66, $\theta$ = 90$^{\circ}$, and $\phi$ = 37$^{\circ}$ as shown by the red lines in  Fig.  \ \ref{fig:spectrum}(a), in excellent agreement with the calculated results.

   We confirmed the directions of EFG at Al(1)  by measuring the $^{27}$Al NMR spectra under different $H$ directions. 
     When $H$ is applied in the $ab$ plane and is perpendicular to the $a$ axis, one can expect that the Al(1) NMR lines will be split into two sets of lines corresponding to the angles ($\theta,\phi$) = (0$^{\circ}$, 0$^{\circ}$) and (60$^{\circ}$, 53$^{\circ}$) due to the relative angles between $H$ and the $V_{\rm ZZ}$ axis for three inequivalent Al(1) sites as shown Fig.  \ \ref{fig:spectrum}(d).
     In fact, as shown in Fig.  \ \ref{fig:spectrum}(b), the observed spectra under $H \perp a$ (and also $H \perp c$) are well explained by the two sets of NMR lines shown by the red and blue lines from Al(1) with  ($\theta$, $\phi$) = (0$^{\circ}$, 0$^{\circ}$) and  (60$^{\circ}$, 53$^{\circ}$), respectively, with the same values of  $\nu_{\rm Q}$~=1.55(2)~MHz, $\eta$ = 0.66 as used in Fig.  \ \ref{fig:spectrum}(a). 
       In addition, the Al(2) lines are also well reproduced by the spectrum calculations shown in green lines with $\theta$ = 90$^{\circ}$  with the same values of $\nu_{\rm Q}$~= 1.07(2)~MHz, $\eta$ = 0.

     For further confirmation,  we measured the $^{27}$Al NMR spectrum under   $H \parallel a$ (and also $H \perp c$), as shown Fig.  \ \ref{fig:spectrum}(c). 
   In this case, one can expect the two sets of NMR spectra for Al(1) due to two different values of $\theta$ =  90$^{\circ}$ and 30$^{\circ}$ [see Fig. \ \ref{fig:spectrum}(d)]. 
   Although we found that the $H$ direction is deviated by 2$^{\circ}$ from the $a$ axis in our experiment, by taking the misorientation into consideration, the observed spectra are reasonably well reproduced by the calculated results shown by the red, orange, and blue lines with different values of ($\theta$, $\phi$)  = (92$^{\circ}$, 53$^{\circ}$), (32$^{\circ}$, 53$^{\circ}$) and (28$^{\circ}$, 53$^{\circ}$), respectively. 
     Note that the green lines are for Al(2) with the exact same parameters in the case of $H \perp a$ (and also $H \perp c$) [Fig.  \ \ref{fig:spectrum}(b)].
  It is noted that we calculated the spectra by fully diagonalizing the nuclear spin Hamiltonian [Eq. (\ref{eq:1})] to ensure that the effects of quadrupolar interaction and Knight shift were fully accounted for in the determination of the positions of NMR lines.

  Figure \ \ref{fig:Shift} shows the $T$ dependence of the Knight shifts for $H\parallel c$ ($K_c$)  and $H \parallel ab$ plane ($K_{ab}$)  determined by fitting the observed spectra for Al(1).
    Within our experimental uncertainty, we did not see the difference in $K$s between Al(1) and Al(2). 
   Both $K_c$  and $K_{ab}$ decrease slightly with decreasing $T$ from 300 K to $\sim$ 50 K and level off at lower $T$ (note $-K_c$  and $-K_{ab}$ are plotted in  Fig. \ \ref{fig:Shift}). 
   As shown in Fig. \ \ref{fig:Shift},  the $T$ dependence of the Knight shifts is well scaled with $\chi$ above $\sim50$ K where $\chi_{ab}$ was tentatively plotted as $\chi$ is isotropic.
   However,  a clear difference between Knight shift and $\chi$  can seen below $\sim$ 50 K.  
  We also measure $K$ at a lower magnetic field of $\sim$ 1.85 T and the nearly $T$-independent behavior of $K$ at low $T$ has been confirmed.       
   Therefore, the upturn in $\chi$ observed at low $T$ in Fig. \ \ref{fig:Shift} is considered to be non-intrinsic and likely arises from a small number of paramagnetic impurities, disorder, or defects. 
            
\begin{figure}[tb]
\includegraphics[width=8.5cm]{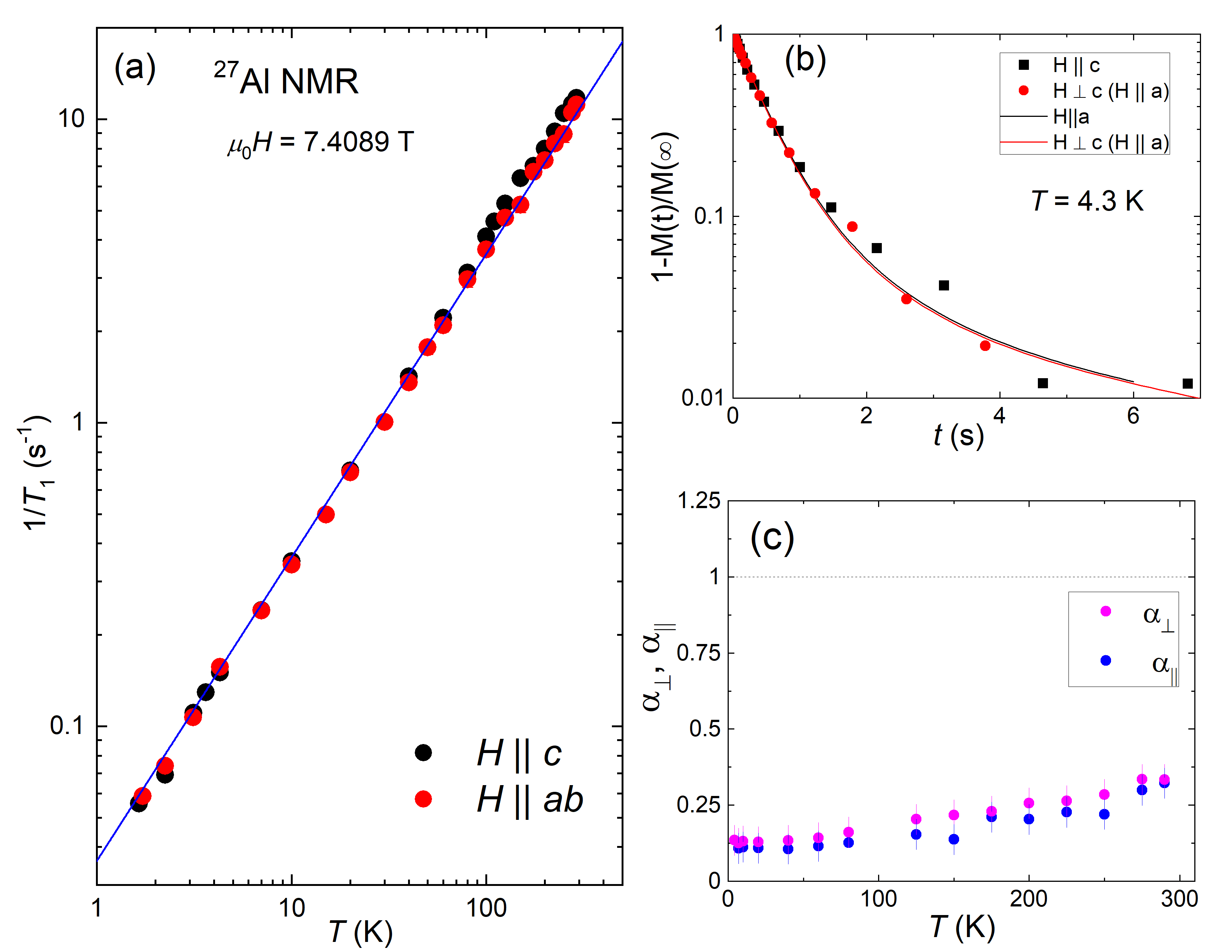} 
\caption{(a) $T$ dependence of $1/T_1$ for both $H$ directions, $H\parallel c$ axis  and  $H\parallel ab$ plane. The solid line shows 1/$T_1$ = 0.036 $T$.  
(b) Recovery curves at $T=4.3$ K for both $H$ directions. The solid lines are fits by the relaxation function described in the text. 
(c) $T$ dependence of $\alpha_\parallel$ and $\alpha_\perp$.}
\label{fig:T1}
\end{figure}

       The Knight shift has contributions from the $T$-dependent spin part $K_{\rm spin}$ and a $T$-independent orbital part $K_0$. 
      $K_{\rm spin}$ is proportional to the spin susceptibility $\chi_{\rm spin}$ through the hyperfine coupling constant $A$ giving $K=K_0+\frac{A}{N_{\rm A}\mu_{\rm B}}\chi_{\rm spin}$, where $N_{\rm A}$ is Avogadro's number.
     The inset of Fig. \ref{fig:Shift}(b)  plots $K_{ab}$ and $K_{c}$ against the corresponding  $\chi_{ab}$  and  $\chi_c$, respectively, with $T$ as an implicit parameter. 
     Here we used the $K_{ab}$ and $K_{c}$ data above 100 K to avoid the contribution to the magnetic susceptibility data originating from the extrinsic contribution observed at low $T$. 
     As shown,  $K_{ab}$ and $K_{c}$ are seen to vary nearly linearly with the corresponding $\chi$, and the hyperfine coupling constants are estimated to be $A_{c}$ = (--1.2 $\pm$ 0.2) T/$\mu_{\rm B}$, and $A_{ab}$ = (--1.2 $\pm$ 0.2) T/$\mu_{\rm B}$ for $H$ $\parallel$ $c$  and $H$ $\parallel$ $ab$, respectively, from the slopes of the linear lines.
      The values of $K_0$ for $H$ $||$ $c$ and  $H$ $\parallel$ $ab$ are estimated to 0.003, and  -0.004 \%, respectively.
   It is noted that our values of the hyperfine coupling constants differ significantly from the previously reported value of 0.0657  T/$\mu_{\rm B}$ both in magnitude and sign \cite{Samanta2024}. 
    Although the direction of $H$ for $^{27}$Al NMR measurements in the previous paper was not specified  \cite{Samanta2024}, their NMR spectra with only one set of $I$ = 5/2 (i.e., five lines) with uneven spacing between lines are quite different from what we observed for at least three different $H$ directions as shown in Figs. \ \ref{fig:spectrum}(a)-(c) in the present study. 
    In addition, no value of $\nu_{\rm Q}$ of Al NMR was reported in their paper \cite{Samanta2024}. 
    At present, although we do not have any clear idea for the quite different NMR spectra, we believe that our NMR spectra are reliable as we observed the two Al NMR signals from the two different Al sites existing in the compound.

    To investigate the dynamical magnetic properties, we have measured 1/$T_1$ versus $T$ (Fig.~\ref{fig:T1}(a)).  
    Here the 1/$T_1$ was measured at the peak position of the spectrum. 
    Since the signal intensity at the peak position is dominated by Al(1), 1/$T_1$ discussed here can be attributed to that of Al(1).   
    $1/T_1$ is nearly isotropic and is roughly proportional to $T$ over the entire $T$ range although a slight deviation seems to be observed at high $T$ above $\sim$ 70 K,  evidencing the metallic state from a microscopic point of view. 
    Fig.~\ref{fig:T1}(b) shows nuclear magnetization recovery curves for the two $H$ directions together with fitting results.   
    As described above, the recovery curves are well fitted with the function with unique $T_1$. This may suggest that the $T_1$ of Al(2) is very similar to that of Al(1).

   To examine the character of the spin fluctuations in detail, we perform a modified Korringa relation analysis. 
Within a Fermi liquid picture, $1/T_1T$ is proportional to the square of the density of states at the Fermi energy ${\cal D}(E_{\rm F})$ and $K_{\text{spin}} (\propto \chi_{\text{spin}}$) is proportional to ${\cal D}(E_{\rm F})$. 
    In particular, $T_1TK_{\text{spin}}^2$  = $\frac{\hbar}{4\pi k_{\rm B}} \left(\frac{\gamma_{\rm e}}{\gamma_{\rm N}}\right)^2$ = $S$, which is the Korringa relation.  
    Deviations from $S$ can reveal information about electron correlations in materials \cite{Moriya1963,Narath1968}, which are expressed via the parameter $\alpha=S/(T_1TK_{\text{spin}}^2)$. 
      For instance, enhancement of $\chi(\mathbf{q}\neq 0)$ increases $1/T_1T$ but has little or no effect on $K_{\text{spin}}$, which probes only the uniform $\chi$ with $\mathbf{q}$ = 0.  
    Thus  $\alpha >1$ for AFM correlations and $\alpha <1$ for FM correlations.

    Since $1/T_{1}T$ probes magnetic fluctuations perpendicular to the magnetic field, it is natural to consider the Korringa ratio $1/T_{1,\bot}TK_{\text{spin},ab}^2$ where $1/T_{1,\bot}T$ = $1/(T_{1}T)_{H\|c}$,  when examining the character of magnetic fluctuations in the $ab$ plane \cite{Wiecki2015-2}. 
   Similarly, we consider the Korringa ratio $1/T_{1,\|}TK_{\text{spin},c}^2$ for magnetic fluctuations along the $c$ axis. 
   Here $1/(T_{1,\|}T)$ is estimated from  $2/(T_{1}T)_{H\|ab}$ $-$ $1/(T_{1}T)_{H\|c}$.

     By utilizing  1/($T_{1, \perp}TK^2_{{\rm spin}, ab}$)  and  1/($T_{1, ||}TK^2_{{\rm spin}, c}$) ,  $\alpha_\perp$ and $\alpha_\parallel$ are calculated whose results are  shown in  Fig.~\ref{fig:T1}(c).
    $\alpha_\parallel$ ($\alpha_\perp$)  decrease from $\sim$0.32 (0.33) at 290 K to $\sim$0.11 (0.13) around 50  K and then become nearly temperature independent below that temperature, indicating dominant ferromagnetic spin correlations at low temperatures.
    Furthermore, $\alpha_\perp$  is close to $\alpha_\parallel$ suggesting that the ferromagnetic correlations are nearly isotropic. 
        The lowest values of  $\alpha_\parallel$ and $\alpha_\perp$  in Sc$_3$Mn$_3$Al$_7$Si$_5$ are slightly greater than nearly ferromagnetic compounds SrCo$_2$P$_2$, BaCo$_2$As$_2$ and SrCo$_2$As$_2$ in which dominant ferromagnetic spin fluctuations have also been reported \cite{ Furukawa2024,Wiecki2015,BaCo2As2}.
       The above analysis is based on a simple model  that the nuclear relaxation is due to the local ${\cal D}$$(E_F)$ at the Al(1) site, through on-site hyperfine interactions, where Al-3$p$ bands hybridize with Mn-3$d$ bands. 
        On the other hand, if the relaxations are induced by only localized Mn electronic spins through isotropic transferred hyperfine interactions, the $\alpha$ value would be modified by a factor of 2 due to the number of the nearest neighbor Mn ions for a given Al(1).
  Regardless of the model, the $\alpha$ values seem to be consistent with FM spin fluctuations.

 \begin{figure}[tb]
\includegraphics[width=0.85\columnwidth]{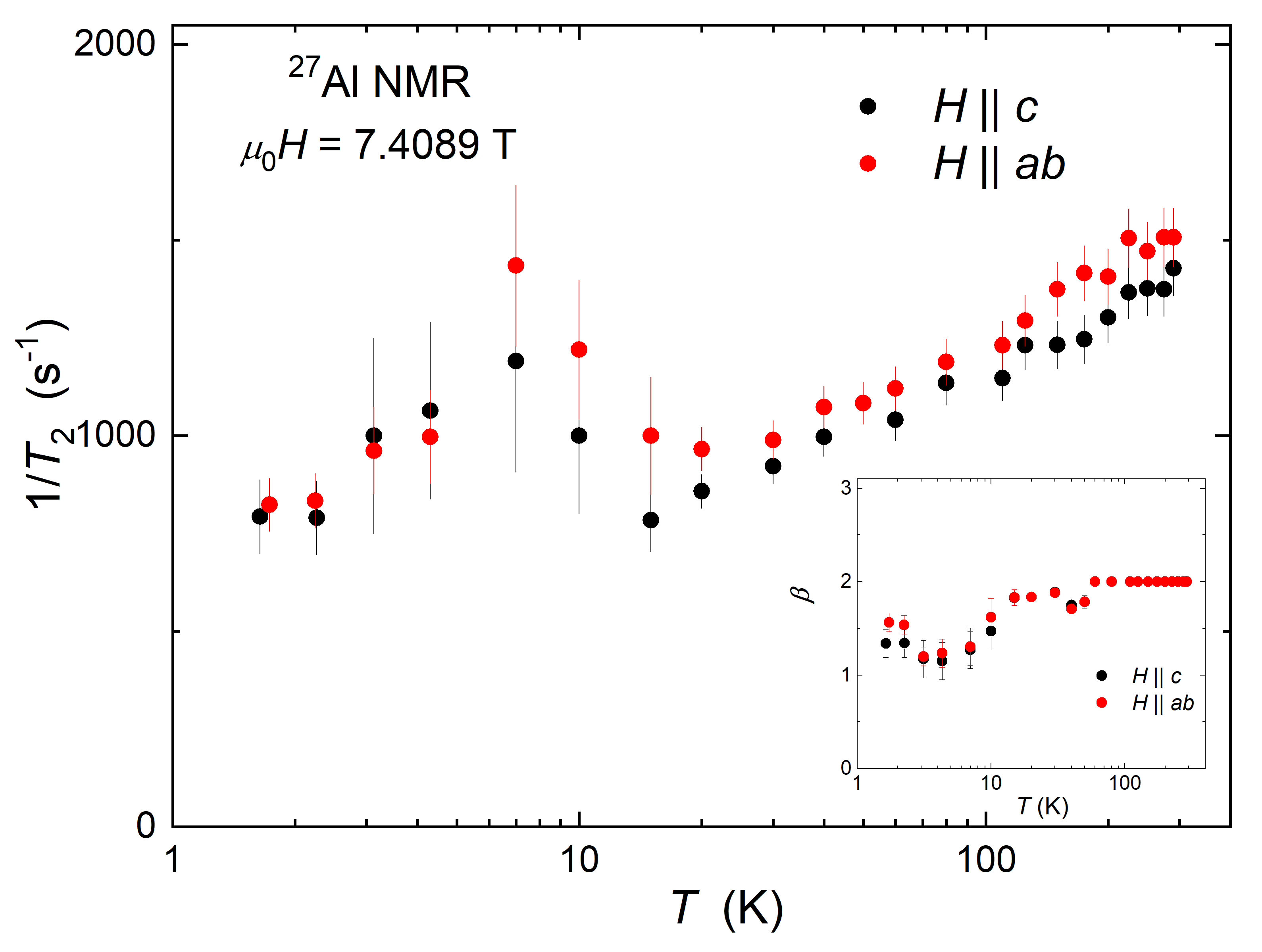} 
\caption{$T$ dependence of $1/T_2$ for both magnetic field directions, $H\parallel c$ axis  and  $H\parallel ab$ plane.
Inset: $T$ dependence of $\beta$ for both $H$ directions, $H\parallel c$ axis  and  $H\parallel ab$ plane. }
\label{fig:T2}
\end{figure}

   Finally, we discuss the slow spin dynamics based on the $T$ dependence of $1/T_2$ measured at the peak position of the $^{27}$Al-NMR spectrum. 
    Here again, we consider that we are measuring $T_2$ at the Al(1) site. 
   As shown in Fig. \ \ref{fig:T2},   $1/T_2$ under both $H\parallel c$ and  $H\parallel ab$ are nearly the same within our experimental uncertainty and exhibit  similar $T$ dependence where, with decreasing $T$,  both 1/$T_2$ gradually decrease down to $T\simeq15-20$~K, and start to increase, and then exhibit peaks around 7 K.

    In general, 1/$T_2$ is related to 1/$T_1$ and can be written  as \cite{Abragambook}
$\frac{1}{T_2 } = \left(\frac{1}{T_2}\right)^* +  \frac{1}{2}F_\perp(\omega_{\rm N}) +F_z(0),$
with 
$ F_\perp(\omega_{\rm N}) =  \frac{1}{T_1 }$. 
Here, $F_\alpha(\omega)$ is the spectral density of the longitudinal ($\alpha$ = $z$) and transverse ($\alpha$ = $\perp$) components of the fluctuating local field ($h_{\alpha}$),  and is described by 
 $ F_\alpha (\omega_{\rm N})   
 =  \frac{1}{2} \gamma_{\rm N}^2 \int_{-\infty}^{+\infty} \langle h_{\alpha}(t) h_{\alpha}(0) \rangle  {\rm exp}(i \omega_{\rm N}t) dt$.
$(1/T_2)^*$ is due to the nuclear dipole-dipole interaction \cite{Slichter_book, Abragambook} and is independent of $T$.
Since the observed 1/$T_2$ shows the $T$ dependence as described above, $(1/T_2)^*$  is not relevant here. 

Thus the $T$-dependent part of 1/$T_2$ comes from the other two terms originating from $F_z(0)$ and $F_\perp(\omega_{\rm N})$.
$F_z(0)$ is driven by the longitudinal component of magnetic fluctuations at nearly zero frequency of the order of kHz, while  $F_\perp(\omega_{\rm N})$ (=1/$T_1$) originates from the transverse components of the fluctuations at the NMR frequency of the order of MHz. 
    As shown above, there is no anomaly around 7 K in the $T$ dependence of 1/$T_1$,  the enhancement observed in 1/$T_2$ can be attributed to   $F_z(0)$.
 This indicates the very slow longitudinal fluctuations of the hyperfine field at the Al(1) site with frequencies in kilohertz or lower frequency along the external field directions.
    Since we observed the FM spin fluctuations below 50 K, the very slow longitudinal hyperfine field fluctuations can be attributed to the FM fluctuations.
Since $F_\perp(\omega_{\rm N})$, that is, 1/$T_1$, picks up the hyperfine fluctuations at NMR frequency of $\omega_{\rm N}$ while $F_z(0)$ relates to the hyperfine fluctuations at $\omega$ = 0, those results suggest that the spectrum density of the FM spin fluctuation extend to $\omega$ $\sim$  0 at low $T$.
  Therefore, we suggest that the FM spin fluctuations slow down at low $T$,  which could be consistent with the previous report suggesting an FM instability due to the flat bands near Fermi Energy \cite{Samanta2024}.

      Recently, magnetic fluctuations have been proposed as the mechanism for the large topological Hall effect at elevated temperature in the kagome metal YMn$_6$Sn$_6$ \cite{Ghimire2020}.
    In the kagome metal LaRu$_3$Si$_2$ superconductor, electronic correlations and ferromagnetic fluctuations are also found to be crucial to understanding the non-Fermi-liquid behavior and the high superconducting transition temperature in this material \cite{Wang2023}.
   The ferromagnetic fluctuations found in this study may play significant roles in the interesting physical properties in Sc$_3$Mn$_3$Al$_7$Si$_5$, which is worthy of further study.

  In conclusion, we conducted $^{27}$Al NMR measurements on  Sc$_3$Mn$_3$Al$_7$Si$_5$  to investigate its static and dynamic magnetic properties. 
       Two distinct $^{27}$Al NMR  signals with two different values of quadrupolar frequencies of $\nu_{\rm Q}$ = 1.55(2) and  1.07(2) MHz are observed, which are assigned to Al(1) and Al(2), respectively. 
      From the detailed NMR spectrum measurements under three different magnetic field directions and the DFT calculations, the principal axes of EFG for each Al site have been determined.  
          Based on a Korringa ratio analysis using the $T_1$ and  $K$  data,  ferromagnetic spin fluctuations are found to dominate in the kagome metal Sc$_3$Mn$_3$Al$_7$Si$_5$ at low temperatures.
    Furthermore, the ferromagnetic fluctuations are suggested to be very slow with frequency of the order of kilohertz or less at low temperatures. 
    Our findings strongly call for further detailed investigations on the nearly ferromagnetic kagome metal Sc$_3$Mn$_3$Al$_7$Si$_5$ to characterize the peculiar magnetic properties.

 \section{Acknowledgments}
  The research was supported by the U.S. Department of Energy, Office of Basic Energy Sciences, Division of Materials Sciences and Engineering. Ames National Laboratory is operated for the U.S. Department of Energy by Iowa State University under Contract No.~DE-AC02-07CH11358.  
C.D. and Y.N. were supported by an NSF Career DMR-1944975.
M.H.P acknowledges support from the U.S. Department of Energy,  Office of Basic Energy Sciences, Division of Materials Science
 and Engineering under grant no. DE-FG02-07ER46438.


\begin{thebibliography}{99}



\bibitem{Yin2022} J.-X. Yin, B. Lian, and M. Zahid Hasan, Topological kagome magnets and superconductors, Nature {\bf 612}, 647 (2022).
\bibitem{Xu2023} X. Xu, J.-X. Yin, Z. Qu, and S. Jia, Quantum interactions in topological R166 kagome magnet,  Rep. Prog. Phys. {\bf 86} 114502 (2023).
\bibitem{Yin2020} J.-X. Yin $et$ $al$., Quantum-limit Chern topological magnetism in TbMn$_6$Sn$_6$, Nature  {\bf 583}, 533 (2020).
\bibitem{Riberolles2022} S. X. M. Riberolles, T. J. Slade, D. L. Abernathy, G. E. Granroth, B. Li, Y. Lee, P. C. Canfield, B. G. Ueland, L. Ke, and R. J. McQueeney, Low-temperature competing magnetic energy scales in the topological ferrimagnet TbMn$_6$Sn$_6$, Phys. Rev. X {\bf 12}, 021043 (2022).
\bibitem{Riberolles2023} S. X. M. Riberolles, T. J. Slade, R. L. Dally, P. M. Sarte, B. Li, T. Han, H. Lane, C. Stock, H. Bhandari, N. J. Ghimire, D. L. Abernathy, P. C. Canfield, J. W. Lynn, B. G. Ueland, and R. J. McQueeney, Orbital character of the spin-reorientation transition in TbMn$_6$Sn$_6$, Nat. Commun. {\bf 14}, 2658 (2023).
\bibitem{Riberolles2024} S. X. M. Riberolles, T. J. Slade, T. Han, B. Li, D. L. Abernathy, P. C. Canfield, B. G. Ueland, P. P. Orth, L. Ke, and R. J. McQueeney, Chiral and flat-band magnetic quasiparticles in ferromagnetic and metallic kagome layers, Nat. Commun.{\bf 15}, 192 (2024).
\bibitem{Cao2023} S. Cao, C. Xu, H. Fukui, T. Manjo, Y. Dong, M. Shi, Y. Liu, C. Cao, and Y. Song, Competing charge-density wave instabilities in the kagome metal ScV$_6$Sn$_6$,  Nat. Commun. {\bf 14}, 7671 (2023). 
\bibitem{Pokharel2021} G. Pokharel, S. M. L. Teicher, B. R. Ortiz, P. M. Sarte, G. Wu, S. Peng, J. He, R. Seshadri, and S. D. Wilson, Electronic properties of the topological kagome metals YV$_6$Sn$_6$ and GdV$_6$Sn$_6$, Phys. Rev. B {\bf 104}, 235139 (2021).

\bibitem{Ortiz2019} Brenden R. Ortiz, Lídia C. Gomes, Jennifer R. Morey, Michal Winiarski, Mitchell Bordelon, John S. Mangum, Iain W.H. Oswald, Jose A. Rodriguez-Rivera, James R. Neilson, Stephen D. Wilson, Elif Ertekin, Tyrel M. McQueen, and Eric S. Toberer, New kagome prototype materials: Discovery of KV3Sb5,RbV3Sb5, and CsV3Sb5, Phys. Rev. Mater. 3, 094407 (2019).
\bibitem{Ortiz2020} R. Ortiz, S. M. L. Teicher, Y. Hu, J. L. Zuo, P. M. Sarte, E. C. Schueller, A. M. M. Abeykoon, M. J. Krogstad, S. Rosenkranz, R. Osborn, R. Seshadri, L. Balents, J. He, and S. D. Wilson, CsV$_3$Sb$_5$: A Z2 Topological Kagome Metal with a Superconducting Ground State, Phys. Rev. Lett.  {\bf 125}, 247002 (2020). 


\bibitem{Wilson2024} S. D. Wilson, and B. R. Ortiz, $A$V$_3$Sb$_5$ kagome superconductors, Nat. Rev. Mater. {\bf 9}, 420 (2024).


\bibitem{Samanta2024} S. Samanta, H. Park, C. Lee, S. Jeon, H. Cui, Y.-X. Yao, J. Hwang, K.-Y. Choi, and H.-S. Kim, Emergence of flat bands and ferromagnetic fluctuations via orbital-selective electron correlations in Mn-based kagome metal, Nat. Commun. {\bf 15} 5376 (2024).
\bibitem{He2014} H. He, W. Miiller, and M. C. Aronson, New Kagome Metal Sc$_3$Mn$_3$Al$_7$Si$_5$ and Its Gallium-Doped Analogues: Synthesis, Crystal Structure, and Physical Properties, Inorg. Chem. {\bf 53}, 9115 (2014).
\bibitem{Li2021} X. Y. Li, D. Reig-i-Plessis, P.-F. Liu, S. Wu, B.-T. Wang,  A. M. Hallas, M. B. Stone, C. Broholm, and M. C. Aronson, Neutron scattering study of the kagome metal Sc$_3$Mn$_3$Al$_7$Si$_5$,  Phys. Rev. B {\bf 104}, 134305 (2021).

\bibitem{Recovery} A. Narath, Nuclear Spin-Lattice Relaxation in Hexagonal Transition Metals: Titanium,  Phys. Rev. {\bf 162}, 320 (1967).
\bibitem{Slichter_book} C. P. Slichter, $Principles~of~Magnetic~Resonance$, 3rd ed. (Springer, New York, 1990).
\bibitem{Abragambook} A. Abragam, $The~Principles~of~Nuclear~Magnetism$. (Clarendon Press, Oxford,1961).
\bibitem{FLAPW}
P. Blaha, K. Schwarz, G. K. H. Madsen, D. Kvasnick, and J. Luitz, WIEN2K, An Augmented Plane Wave+Local Orbitals Program for Calculation Crystal Properties (Karlheinz Schwarz, Technical Universität Wien, 2001).
\bibitem{GGA} J. P. Perdew, K. Burke, and M. Ernzerhof, Phys. Rev. Lett. {\bf 77}, 3865 (1996).
\bibitem{PreviouNMR}  The observed $^{27}$Al NMR spectra are quite different from the previous report in \cite{Samanta2024}.
 \bibitem{Moriya1963}T. Moriya, The Effect of Electron-Electron Interaction on the Nuclear Spin Relaxation in Metals, J. Phys. Soc. Jpn. {\bf 18}, 516 (1963).
\bibitem{Narath1968}A. Narath and H. T. Weaver, Effects of Electron-Electron Interactions on Nuclear Spin-Lattice Relaxation Rates and Knight Shifts in Alkali and Noble Metals, Phys. Rev. {\bf 175}, 378 (1968).
 \bibitem{Wiecki2015-2}  P.  Wiecki, B. Roy, D. C. Johnston, S. L. Bud'ko, P. C. Canfield, and Y. Furukawa, Competing Magnetic Fluctuations in Iron Pnictide Superconductors: Role of Ferromagnetic Spin Correlations Revealed by NMR, Phys. Rev. Lett. {\bf 115}, 137001 (2015).
\bibitem{Furukawa2024} N. Furukawa, Q.-P. Ding, J. Schmidt,  S. L. Bud'ko, P.  C. Canfield, and Y. Furukawa,  Inhomogeneous magnetic ordered state and evolution of magnetic fluctuations in Sr(Co$_{1-x}$Ni$_x$)$_2$P$_2$revealed by $^{31}$P NMR,  Phys. Rev. B {\bf 110}, 014439 (2024).
\bibitem{Wiecki2015} P. Wiecki, V. Ogloblichev, A. Pandey, D. C. Johnston,  and Y. Furukawa,  Coexistence of antiferromagnetic and ferromagnetic spin correlations in SrCo$_2$As$_2$ revealed by $^{59}$Co and $^{75}$As NMR, Phys. Rev. B {\bf 91}, 220406(R) (2015).
\bibitem{BaCo2As2}  K. Ahilan, T. Imai, A. S. Sefat, and F. L. Ning, NMR investigation of spin correlations in BaCo$_2$As$_2$,  Phys. Rev. B {\bf 90}, 014520 (2014).
\bibitem{Ghimire2020} N. J. Ghimire, R. L. Dally, L. Poudel, D. C. Jones, D. Michel, N. T. Magar, M. Bleuel, M. A. McGuire, J. S. Jiang, J. F. Mitchell, J. W. Lynn, and I. I. Mazin, Competing magnetic phases and fluctuation-driven scalar spin chirality in the kagome metal YMn$_6$Sn$_6$, Sci. Adv. {\bf 6}, eabe2680 (2020).
\bibitem{Wang2023} Y. Wang,  Electronic correlation effects on stabilizing a perfect Kagome lattice and ferromagnetic fluctuation in LaRu$_3$Si$_2$, JUSTC {\bf 53}, 0702 (2023).
    \end{thebibliography}
\end{document}